\documentclass{PoS}

\usepackage{caption}
\usepackage{subcaption}
\title{$B$, $B_s$, $K$ and $\pi$ weak matrix elements with physical light quarks }

\ShortTitle{$B$, $B_s$, $K$ and $\pi$  weak matrix elements with physical light quarks}

\author{\speaker{R.~J.~Dowdall}$^a$, C.~T.~H.~Davies$^b$, R.~R.~Horgan$^a$, G.~P.~Lepage$^c$, C.~McNeile$^d$, C.~J.~Monahan$^e$, J.~Shigemitsu$^f$\\
HPQCD collaboration\\
        \llap{$^a$}DAMTP, University of Cambridge, Wilberforce Road, Cambridge CB3 0WA, UK\\
        \llap{$^b$}SUPA, School of Physics and Astronomy, University of Glasgow, Glasgow, G12 8QQ, UK\\
        \llap{$^c$}Laboratory of Elementary-Particle Physics, Cornell University, Ithaca, New York 14853, USA\\
        \llap{$^d$}Bergische Universit\"{a}t Wuppertal, Gaussstr.\,20, D-42119 Wuppertal, Germany\\
        \llap{$^e$}Physics Department, College of William and Mary, Williamsburg, Virginia 23187, USA\\
        \llap{$^f$}Physics Department, The Ohio State University, Columbus, Ohio 43210, USA\\
        E-mail: \email{r.j.dowdall@damtp.cam.ac.uk} }

\abstract{Calculations of pseudoscalar decay constants of $B$, $B_s$, $K$ and $\pi$ mesons with physical light quarks are presented. We use HISQ ensembles that include $u,d,s$ and $c$ sea quarks at three lattice spacings.
HISQ is used for the valence light quarks and a radiatively improved NRQCD action for the heavy quarks.  
The key results are $f_{B^+}=0.184(4)$ GeV, $f_{B_s}=0.224(4)$ GeV, $f_{B_s}/f_{B^+}=1.217(8)$, $f_{K^+}/f_{\pi^+}=1.1916(21)$, $f_{K^+}=155.37(34)$ MeV, giving a significant improvement over previous results that required chiral extrapolation.
We also calculate the Wilson flow scale $w_0$, finding $w_0=0.1715(9)$ fm.
}

\FullConference{31st International Symposium on Lattice Field Theory - LATTICE 2013\\
		July 29 - August 3, 2013\\
		Mainz, Germany}

\begin{document}

\section{$N_f=2+1+1$ HISQ ensembles inlcuding physical light quarks}
Lattice QCD calculations of decay constants containing light quarks have historically suffered from large uncertainties due to extrapolations to the correct pion mass. Since precision calculations of pseudoscalar decay constants are a central goal of lattice flavour physics, finding use in predicting rare decays and determining standard model parameters, it is a worthwhile investment to compute directly at the physical point. A number of collaborations are now generating ensembles with physical light quark masses, including the MILC collaboration \cite{Bazavov:2010ru,Bazavov:2012xda} who use a Symanzik improved gluon action and include $N_f=2+1+1$ flavours of sea quarks with the highly improved staggered quark (HISQ) action.
We employ eight gluon ensembles with scales from $0.09-0.15$ fm and with pion masses ranging from $330$ MeV down to physical, the details are shown below:
\begin{center}
\begin{tabular}{llllllllll}
\hline
Set & $\beta$ & $a$ (fm) 	& $M_\pi$ (MeV) & L (fm)    & $L/a \times T/a$ & $n_{{\rm cfg}}$  \\
\hline
1 & 5.8 & 0.15  & 300 & 2.5 & 16$\times$48 & 1020 \\
2 & 5.8 & 0.15  & 215 & 3.7 & 24$\times$48 & 1000 \\
3 & 5.8 & 0.15  & 130 & 4.8 & 32$\times$48 & 1000 \\
\hline
4 & 6.0 & 0.12   & 300 & 3.0 & 24$\times$64 & 1052 \\
5 & 6.0 & 0.12   & 215 & 3.9 & 32$\times$64 & 1000 \\
6 & 6.0 & 0.12   & 130 & 5.8 & 48$\times$64 & 1000 \\
\hline
7 & 6.3 & 0.09   & 300 & 2.9 & 32$\times$96 & 1008 \\
8 & 6.3 & 0.09   & 130 & 5.6 & 64$\times$96 & 621 \\ \hline
\end{tabular}
\end{center}
HPQCD have previously studied the $\Upsilon$ and B-meson spectra on a subset of these ensembles \cite{Dowdall:2011wh,Dowdall:2012ab,Daldrop:2011aa}.

\begin{figure}
\centering
\begin{minipage}{.49\textwidth}
  \centering
  \includegraphics[width=0.99\hsize]{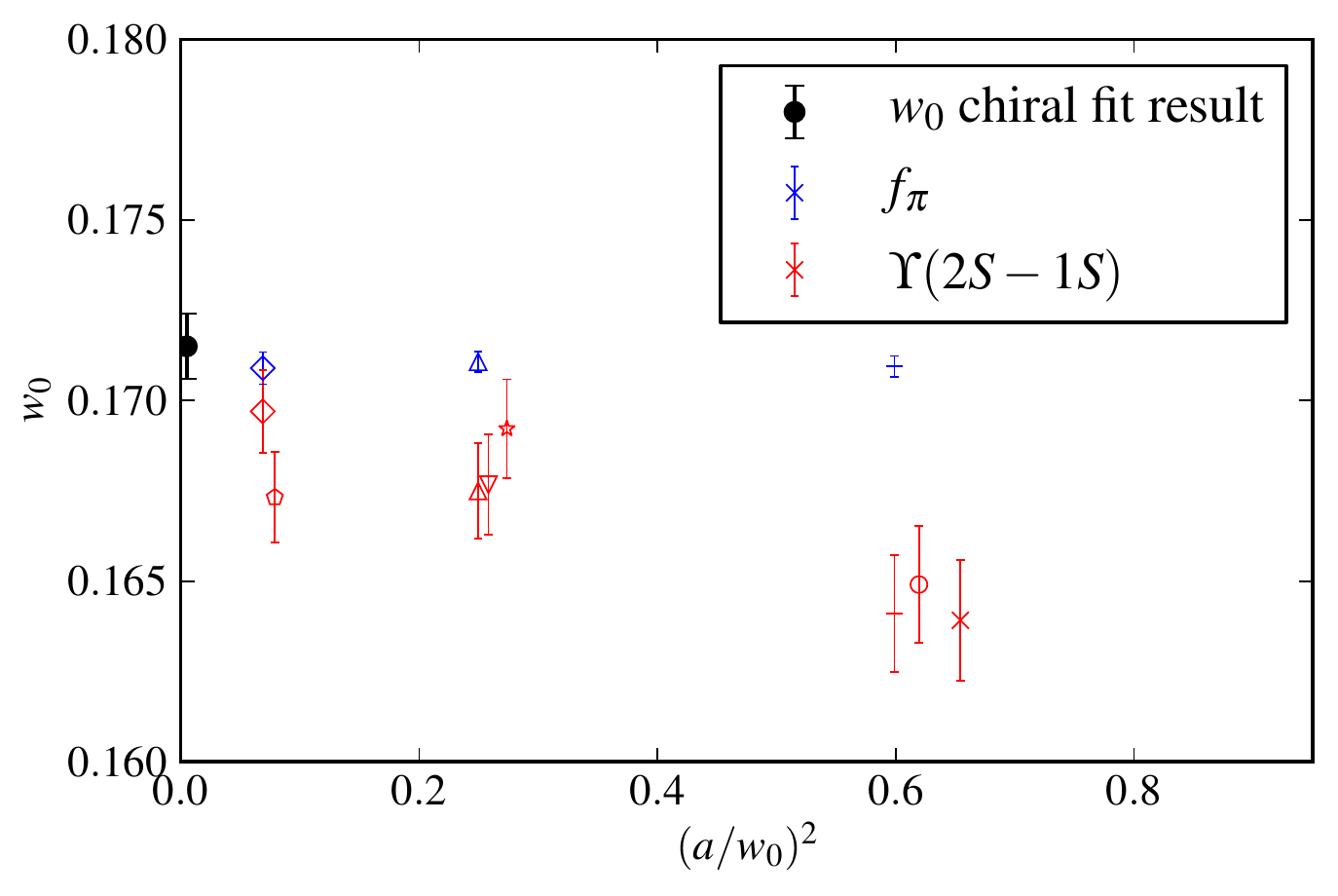}
  \captionof{figure}{
   Comparison of the value of $w_0$ obtained setting the scale using $f_\pi$ (statistical errors only) or the $\Upsilon(2S-1S)$ splitting (including NRQCD systematic error).  
   Both agree as $a\rightarrow 0$. 
   }
  \label{fig:w0a}
\end{minipage}%
\hfill
\begin{minipage}{.49\textwidth}
  \centering
  \includegraphics[width=0.99\hsize]{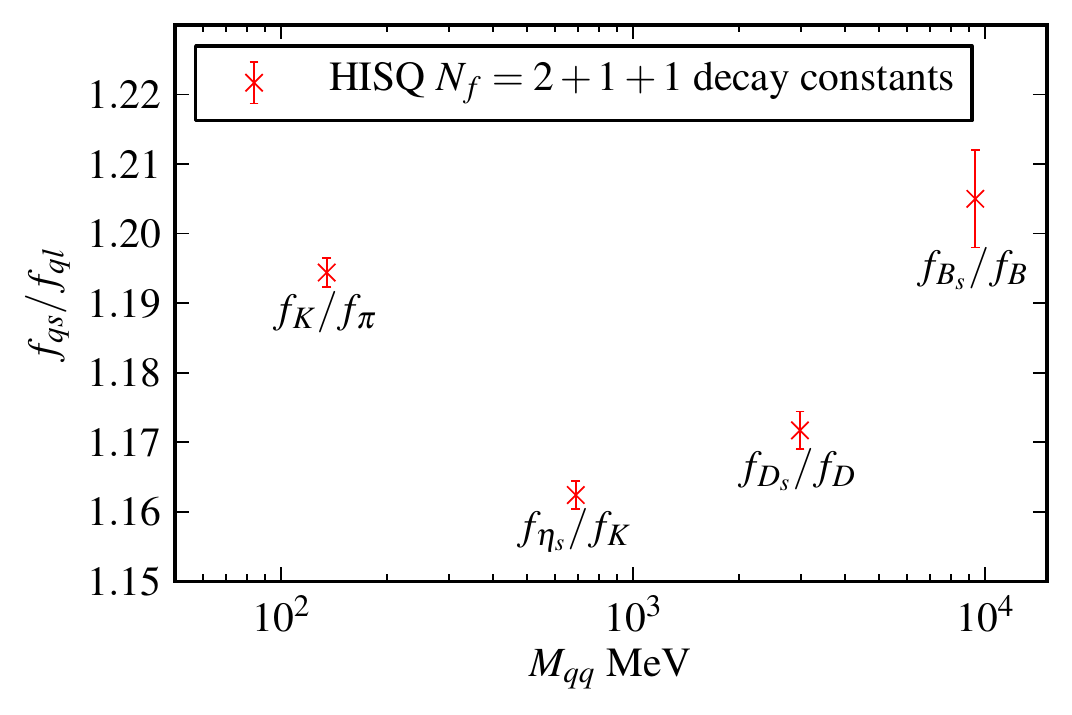}
  \captionof{figure}{Plot of $SU(3)$ breaking decay constant ratios showing the dependence on the ``spectator'' quark mass. $M_{qq}$ denotes the meson made of two quarks $q$. $f_{qs},f_{ql}$ are the decay constants of the mesons with $q$ and either a strange or light quark. }
  \label{fig:spectators}
\end{minipage}
\end{figure}

We have calculated the lattice spacing of each ensemble using a variety of methods, the $\Upsilon(2S-1S)$ splitting, $r_1/a$ from MILC and the Wilson flow scales $w_0, \sqrt{t_0}$ \cite{Luscher:2010iy}. We have used the $\Upsilon(2S-1S)$ for the heavy decay constants results and $w_0$ for the light decay constants, using $f_{\pi^+}$ to set the overall scale.
$w_0/a$ was determined with the Wilson (as opposed to Symanzik improved) flow as in \cite{Borsanyi:2012zs} after binning over 12 adjacent configurations (60-72 MD steps.)
Fig.~\ref{fig:w0a} compares determinations of $w_0$ using $f_\pi$ and the $\Upsilon(2S-1S)$, which agree as $a\rightarrow0$ albeit with larger errors and scaling violations for the $\Upsilon$ method.

These proceedings summarise the results of two papers \cite{Dowdall:2013tga,Dowdall:2013rya} to which we refer the reader for more details.

\section{$B$, $B_s$ decay constants}

The $B$ meson decay constants are calculated using a $v^4$ NRQCD action including 1-loop radiative corrections to most of the Wilson coefficients \cite{Dowdall:2011wh}. In particular, it includes corrections to the spin-magnetic coupling that generates the hyperfine splittings. This was shown to give excellent agreement with experimental splittings in both the B-meson \cite{Dowdall:2012ab} and bottomonium \cite{Dowdall:2011wh} spectra. The NRQCD currents are perturbatively matched to the full QCD current $\langle A_0 \rangle$ through $\mathcal{O}\left(  \alpha_s, \alpha_s \Lambda_{\rm QCD}/m_b  \right)$ using
\begin{equation}
J_0^{(0)} = \bar{\Psi}_q \gamma_5 \gamma_0 \Psi_Q , \ \ \ \ \ 
J_0^{(1)} = \frac{-1}{2m_b} \bar{\Psi}_q \gamma_5 \gamma_0 \gamma \cdot \nabla \Psi_Q ,\ \ \ \ \ 
J_0^{(2)} =\frac{-1}{2m_b} \bar{\Psi}_q \gamma \cdot \overleftarrow{\nabla}   \gamma_5 \gamma_0 \Psi_Q , 
\end{equation}
and the relation
\begin{equation}
  \langle A_0 \rangle =
 (1+ \alpha_s z_0) \left(  \langle J_0^{(0)} \rangle
 +  (1+ \alpha_s z_1  ) \langle J_0^{(1)} \rangle 
  +   \alpha_s  z_2   \langle J_0^{(2)} \rangle
\right).
\end{equation}
Where we used $\alpha_V$ at $q=2/a$.
We find the overall renormalisation to be very small at 0.8\% on the 0.15 fm lattice and 0.2\% on the 0.09 fm lattice \cite{Dowdall:2013tga,Monahan:2012dq}. We allow an error for missing higher 
order effects by allowing for an $\alpha_s^2$ coefficient which is 10 times the order $\alpha_s$ coefficient. 

32 random wall sources were used on each configuration including two different gaussian smeared sources for the $b$ quark, which are fit together in a multi-exponential Bayesian fit. This was found to give good statistical errors for the ground state energies and matrix elements \cite{Dowdall:2012ab}. We performed two separate analyses: one using NLO heavy meson chiral perturbation theory with discretisation terms on all eight ensembles, the other with only the physical point results. 
The chiral fit is performed simultaneously to $M_{B_s}-M_B,f_B,f_{B_s}$:
 \begin{eqnarray}
  \Phi_s &=& f_{B_s}\sqrt{M_{B_s}} = \Phi_{s0}(1.0 + b_s M_\pi^2/ \Lambda_\chi^2 )\\
  \Phi   &=& f_{B}\sqrt{M_{B}}     = \Phi_{0} \left( 1.0 + b_l \frac{M_\pi^2}{\Lambda_\chi^2} + \frac{1+3g^2}{2 \Lambda_\chi^2}  \left( -\frac{3}{2}
  M_\pi^2 \log(M_\pi^2/\Lambda_\chi^2)
  \right) \right).
 \end{eqnarray}
The coefficients $b_s,b_l$ have priors of $0\pm 1$, and $\Lambda_\chi$ denotes the chiral scale.
Discretisation terms are allowed by multiplying the function by $(1+ d_1 (\Lambda a)^2 + d_2 (\Lambda a)^4 )$ with a scale $\Lambda=0.5$ GeV and priors of $0\pm 1$ on the $d_i$, which are also allowed a mild dependence on $am_b$. We take a prior on $g$ of 0.5(5) which encompasses most recent values.
Finite volume corrections are included via the chiral logarithms and have negligible effect.
The physical point only analysis includes only the $d_i$ terms and sets 3,6,8.
We also compared the results from a fit to staggered HM$\chi$PT \cite{Aubin:2005aq,Aubin:2003uc} which gave consistent results.

The results from both analyses agree within 1-$\sigma$ and we quote the chiral fit results as this allows us to correct for isospin breaking effects that distinguish the $B^+$ and $B^0$ mesons. The effect is a 2 MeV shift from the average u/d mass, which is a 1-$\sigma$ effect.  


\begin{figure}
\centering
\begin{subfigure}{.49\textwidth}
  \centering
  \includegraphics[width=0.99\hsize]{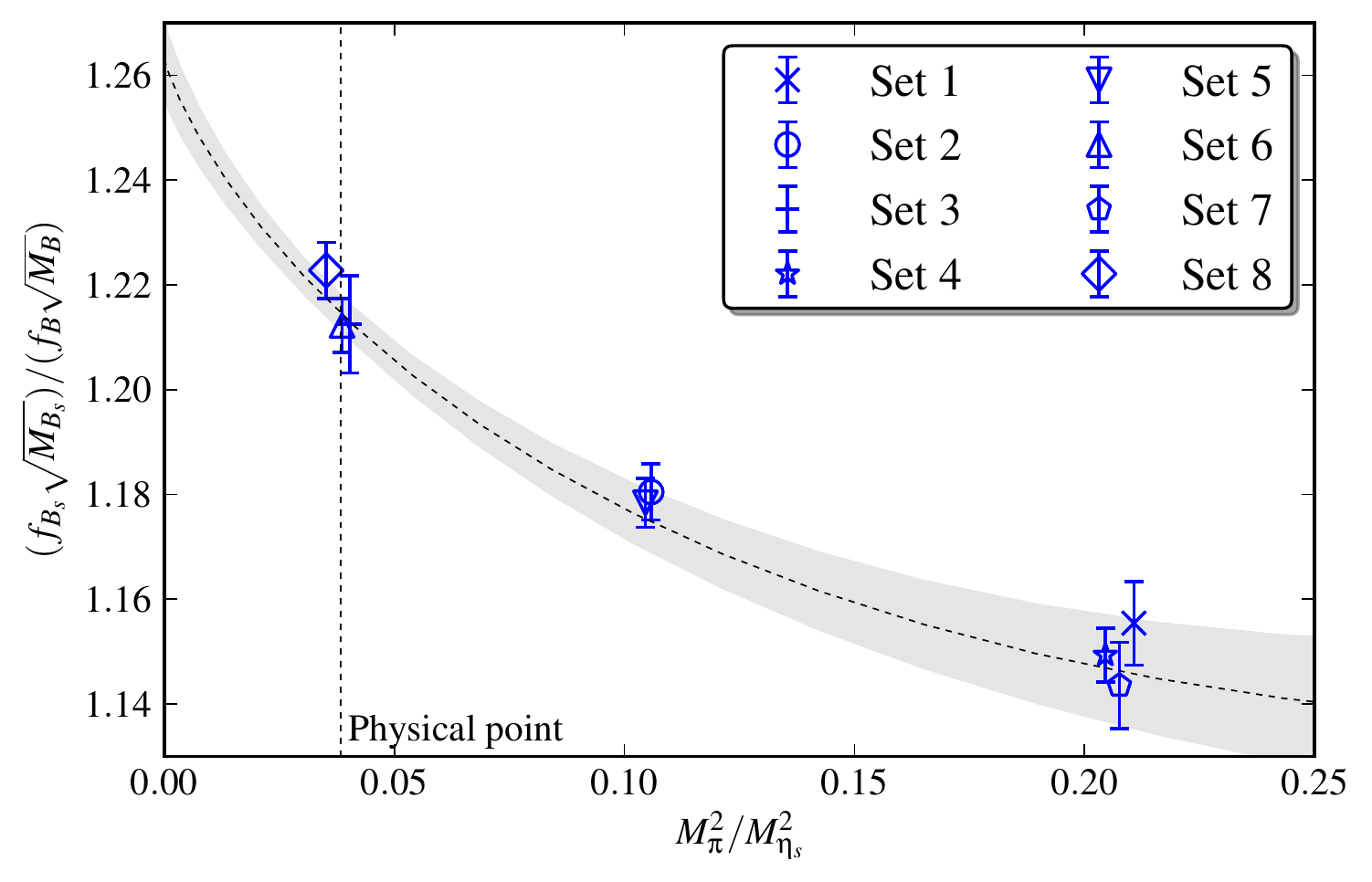}
\end{subfigure}
\hfill
\begin{subfigure}{.495\textwidth}
  \centering
  \includegraphics[width=0.99\hsize]{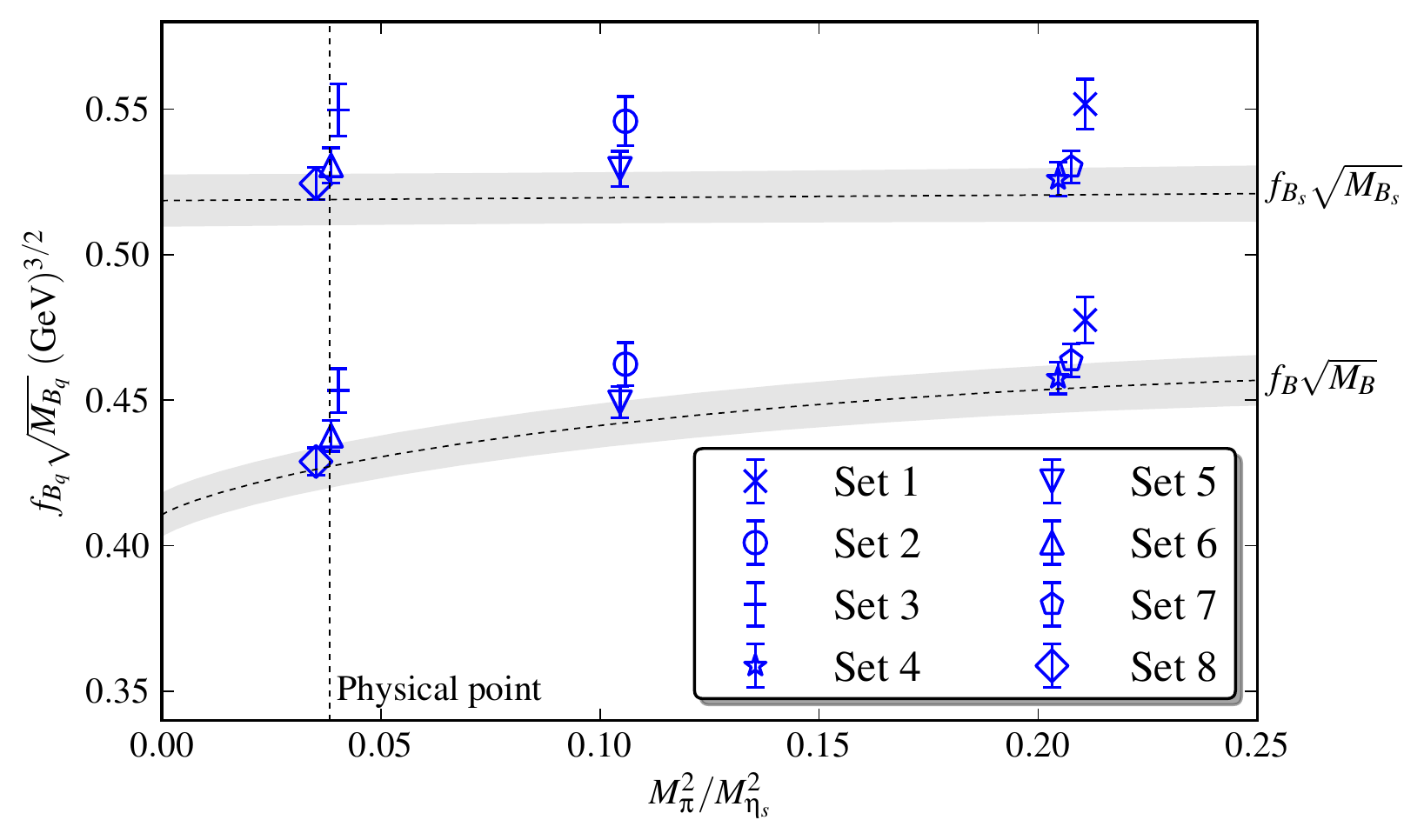}
\end{subfigure}
\caption{Chiral fit to $B$-meson decay constants, and their ratio, against $M_\pi^2/ M_{\eta_s}^2$. Errors on the points are statistical/scale only, errors on the fit bands (grey) include error estimates for missing higher loop renormalisation.}
\label{fig:fb}
\end{figure}


The fit is shown in Fig.~\ref{fig:fb} and a full error budget in \ref{fig:budget1}. The errors are typically dominated by missing higher loop renormalisation or higher order NRQCD operators, whose error we have estimated by power counting. 
The final results are:
 \begin{eqnarray}
f_{B^+} = 0.184(4) {\rm \ GeV} , \ \ \ \ 
f_{B_s} =  0.224(4) {\rm \ GeV}   , \ \ \ \ 
f_{B_s}/f_{B^+} = 1.217(8)  , \ \ \ \ 
M_{B_s} - M_B =  85(2)  {\rm \ MeV }, \nonumber
\end{eqnarray}
where the mass splitting includes an estimated shift of 1(1) MeV for missing electromagnetism and is in good agreement with experiment (87.3(3) MeV \cite{pdg}).

\begin{table}
\centering
\begin{subtable}{.54\textwidth}
  \centering
\begin{tabular}{lcccc}
Error \%	& \small $\Phi_{B_s}/\Phi_B$ & {\small$M_{B_s}-M_B$} & \small$\Phi_{B_s}$ &\small $\Phi_{B}$ \\
\hline
EM: 		& 0.0 	& 1.2 	& 0.0 & 0.0 \\ 
$a$ dep.:	& 0.01 	& 0.9 	& 0.9 & 0.9 \\ 
chiral: 	& 0.01 	& 0.2 	& 0.04 & 0.04\\ 
g: 		& 0.01 	& 0.1 	& 0.0 & 0.01 \\ 
stat/scale: 	& 0.30  & 1.2 	& 0.7 & 0.7 \\ 
operator: 	& 0.0   & 0.0 	& 1.3 & 1.3 \\ 
relativistic: 	& 0.5   & 0.5 	& 1.0 & 1.0 \\ 
\hline
Total: 		& 0.6  & 2.0   & 2.0  & 2.0 \\ 
\hline
\end{tabular}
  \caption{Error budget for the heavy meson decay constants and mass splittings in percent.}
  \label{fig:budget1}
\end{subtable}%
\hfill
\begin{subtable}{.45\textwidth}
  \centering
\begin{tabular}{lllll}
                        & $f_{K^+}$ & $f_{K^+}/f_{\pi^+}$ &  $w_0$ \\ \hline
stats $+$ \emph{svd} cut& 0.13    & 0.13 &  0.26 \\
chiral fit 		&  0.03     & 0.03   &  0.15 \\
$a^2$ extrap 		& 0.10      & 0.10   &  0.27  \\ 
fvol  			& 0.01      & 0.01   &  0.02 \\
$w_0/a$ stats	 	& 0.02      & 0.02   &  0.28 \\
$f_{\pi^+}$ expt	& 0.13      & 0.03   &  0.19 \\
$m_u/m_d$  		& 0.07      & 0.07   &  0.00 \\ \hline
Total 			& 0.22	    & 0.18   &  0.54 \\ \hline 
\end{tabular}
  \caption{Error budget for the light decay constants and $w_0$ in percent.}
  \label{fig:budget2}
\end{subtable}
\caption{Error budgets for the two calculations.}
\label{fig:budget}
\end{table}

\section{$K,\pi$ decay constants }

Decay constants of the $\pi,K$ and fictitious $\eta_s$ ($\bar s s$) meson were calculated from pseudoscalar correlators using the PCAC relation so that no renormalisation is required with the HISQ action. We used 16 $U(1)$ random wall sources on each configuration. The $\pi,K,\eta_s$ are fit simultaneously with 6 exponentials and the correlations between the results are stored for use in the chiral fit.
We begin by multiplying the results by $w_0/a$ to perform the fit in units of $w_0$ which is a free parameter in the fit with prior 0.1755(175). The overall scale is then set by $f_{\pi}^+$.

We then perform a Bayesian fit with $SU(3)$ NLO PQ$\chi$PT \cite{Sharpe:2000bc} supplemented by higher order and discretisation terms. It is worth noting that since we include data at physical pion masses the fit is used just to correct for mistunings of the bare quark masses which are small effects.
The fit function is of the form
\begin{equation}
  f_\mathrm{NLO}(x_a, x_b, x_\ell^\mathrm{sea}, x_s^\mathrm{sea}, L)
  + \delta f_\chi + \delta f_\mathrm{lat},
 \end{equation}
where the $\chi$PT is expressed in terms of the bare meson masses $x_\ell = {{M}_{0,\pi}^2}/{\Lambda_\chi^2},\ x_s = (2 M_{0,K}^2 - M_{0,\pi}^2)/{\Lambda_\chi^2}$.
 Subtracting the 1-loop chiral correction removes (small) finite volume corrections in the masses.
 Generic higher order terms are included in the fit 
\begin{eqnarray}
	\delta f_\chi& \equiv& c_{2a} (x_a+x_b)^2 + c_{2b} (x_a-x_b)^2
		+ c_{2c} (x_a+x_b)(2x_\ell^\mathrm{sea} + x_s^\mathrm{sea})	
		+ c_{2d} (2x_\ell^\mathrm{sea} + x_s^\mathrm{sea})^2
				\nonumber \\
		&&+ c_{2e} (2x_\ell^\mathrm{sea\,2} + x_s^\mathrm{sea\,2}) 
		+ c_{3a} (x_a + x_b)^3 
		+ c_{3b} (x_a+x_b)(x_a-x_b)^2 
		+ c_{3c}(x_a+x_b)^2 (2x_\ell^\mathrm{sea} + x_s^\mathrm{sea}) 
	\nonumber \\
		&&+c_{4} (x_a+x_b)^4 + c_5 (x_a+x_b)^5 + c_6 (x_a+x_b)^6,
	\nonumber
\end{eqnarray}
with priors of 0(1). Discretisation terms are included up to $a^8$ with the coefficients allowed to depend on the meson masses to model taste breaking
\begin{eqnarray}
	\delta f_\mathrm{lat} &\equiv& \sum_{n=1}^4 d_n \left(
	\frac{a\Lambda}{\pi} \right)^{2n}  , \ \ \ \ \
	d_n = d_{n,0} + d_{n,1a} (x_a+x_b) + 
	d_{n,1b} (2x_\ell^\mathrm{sea} + x_s^\mathrm{sea}) 
	 + d_{n,1c} (x_a+x_b)^2. \nonumber
\end{eqnarray}
 Priors of 0(1) were used with the scale set to $ \Lambda = 1.8$ GeV.  This is adequate to account for the discretisation effects in the data and the fit implies a much lower scale. We have also compared a fit with staggered $\chi$PT \cite{Aubin:2005aq,Aubin:2003uc} which agrees within 1-$\sigma$. 
Finite volume effects in the chiral logarithms are computed numerically and found to be less than 0.5\% on all but the smallest ensemble. We allow for higher order finite volume corrections by multiplying this by 1.00(33).

We have compared two methods of estimating the statistical errors in the fit. Firstly, we fit the correlator data binned over 2-4 adjacent configurations. With this approach an SVD cut is needed in the chiral fit due to round-off errors in the covariance matrix which increases the statistical error 
in the chiral fit significantly.
Secondly, we binned over 16 adjacent configurations (80-96 MD steps) in the correlator fits which, with the increased errors, allowed for a fit without the svd cut. The result agrees with the first fit within 1-$\sigma$ and with smaller errors. We quote the first fit, with the larger error.
 
\begin{figure}
\centering
\begin{subfigure}{.5\textwidth}
  \centering
  \includegraphics[width=0.99\hsize]{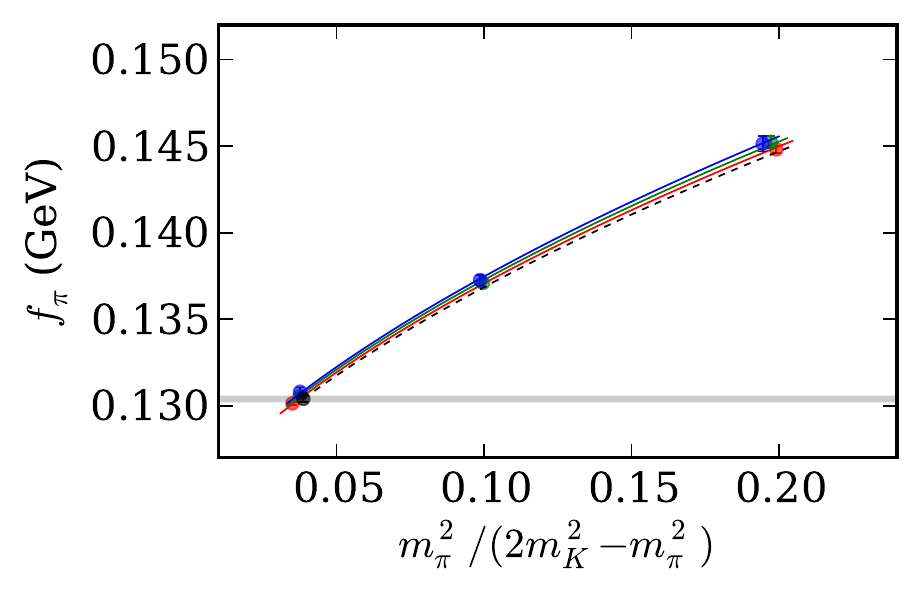}
  \caption{Plot of $f_\pi$ fit.}
  \label{fig:fk2}
\end{subfigure}%
\begin{subfigure}{.5\textwidth}
  \centering
  \includegraphics[width=0.99\hsize]{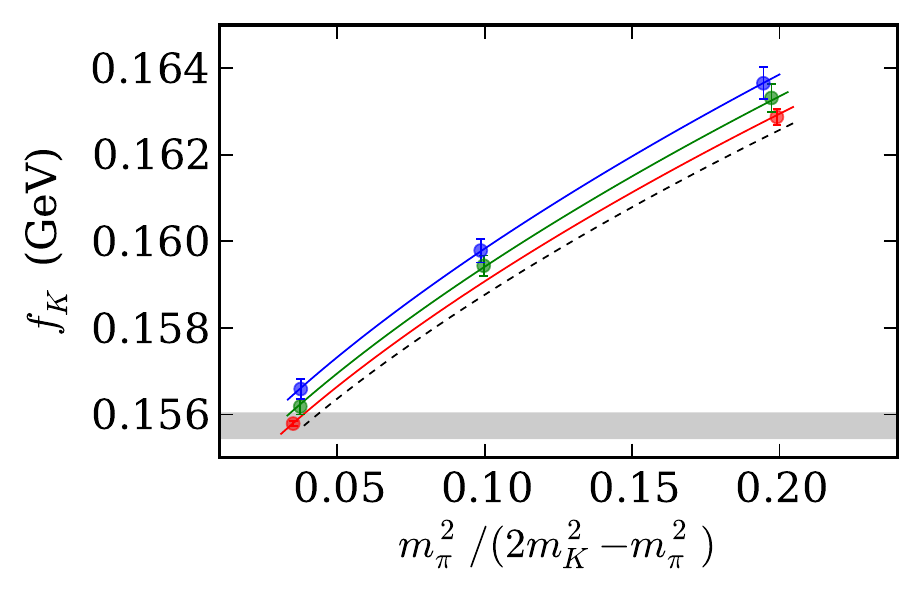}
  \caption{Plot of $f_K$ fit.}
  \label{fig:fk3}
\end{subfigure}
\begin{subfigure}{.5\textwidth}
  \centering
  \includegraphics[width=0.99\hsize]{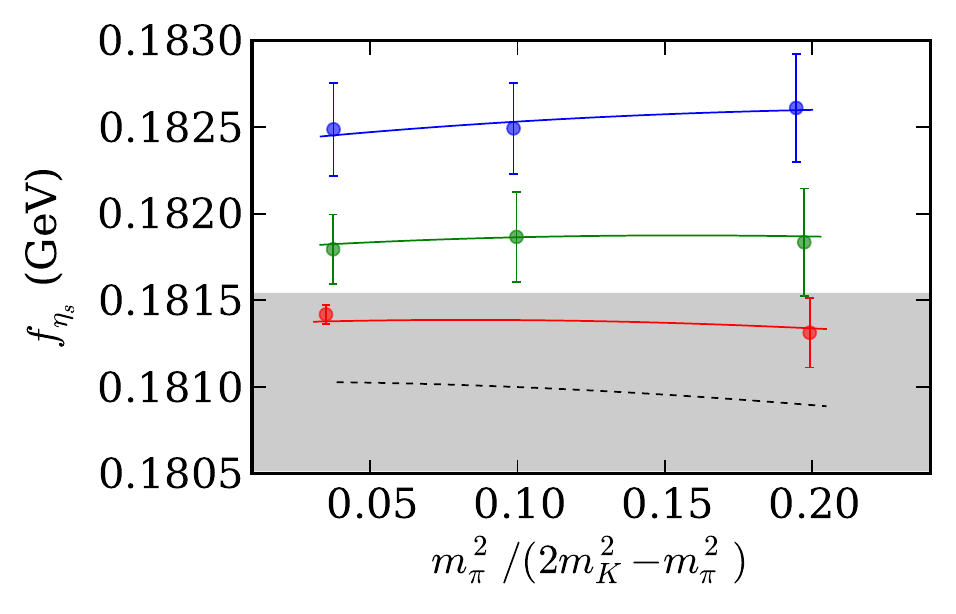}
  \caption{Plot of $f_{\eta_s}$ fit.}
  \label{fig:fk4}
\end{subfigure}
\begin{subfigure}{.5\textwidth}
  \centering
  \includegraphics[width=0.99\hsize]{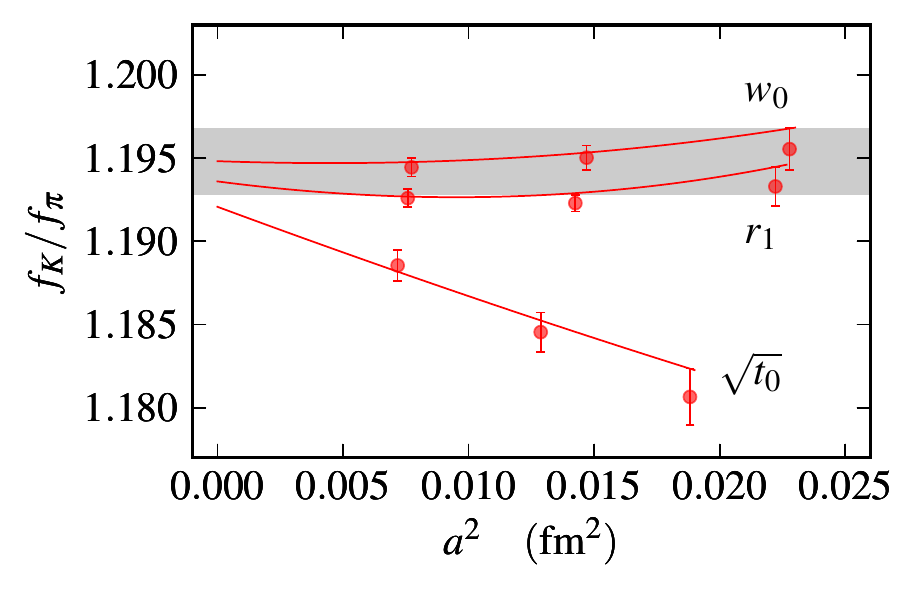}
  \caption{Fit results using $w_0, \sqrt{t_0}$ and $r_1$. The data points have been interpolated to the correct light and strange quark masses.}
  \label{fig:fk1}
\end{subfigure}%
\caption{Chiral and continuum fits to light decay constants. For (a)-(c) the 0.15, 0.12, and 0.09 fm ensembles are in blue, green and red respectively.}
\label{fig:fk}
\end{figure}

The fit is evaluated at the appropriate mass for the  $\pi^+$ and $K^+ $ correcting for isospin and EM effects. This is done using $m_\pi = \sqrt{0.65(9)}M_{\pi}^\mathrm{phys}$ (from the PDG $m_u/m_d$) with $2M_K^2-m_\pi^2$ fixed and by allowing for corrections to Dashen's theorem, see \cite{Dowdall:2013rya} for more detail.
The key results are:
\begin{eqnarray}
	f_{K^+}/f_{\pi^+} = 1.1916(21) \nonumber , \ \ \ \
	f_{K^+} = 155.37(34)\,\mathrm{MeV}  , \ \ \ \
	f_K/f_{K^+}=1.0024(6) \\
	f_{\eta_s} = 181.14(55)\,\mathrm{MeV}  , \ \ \ \ 
	M_{\eta_s} = 688.5(2.2)\,\mathrm{MeV}  , \ \ \ \ 
	w_0 = 0.1715(9)\,\mathrm{fm} 
\end{eqnarray}
The fit results are shown in Figs.~\ref{fig:fk} with the data corrected for strange quark mistuning. The error budget is shown in Fig.~\ref{fig:budget2}. 
We have compared three methods for setting the scale in the fit, $w_0,\sqrt{t_0}$ and $r_1$ all of which agree within a standard deviation. We find that $\sqrt{t_0}$ has worse scaling behaviour than $w_0$ and, combined with the difficulties in determining $r_1/a$, $w_0$ is our preferred method.
Our result for $f_{K^+}/f_{\pi^+}$ gives a determination of $|V_{us}|$ with lattice errors comparable to other sources and an improved test of 1st row unitarity
$$
|V_{us}| = 0.22564(28)_{\mathrm{Br}(K^+)}(20)_{\mathrm{EM}}(40)_{\mathrm{latt}}(5)_{V_{ud}}
, \ \ \ 
1-|V_{ud}|^2-|V_{us}|^2-|V_{ub}|^2 = -0.00009(51). 
$$

\section{Discussion}

We have presented results for pseudoscalar decay constants using physical light quark masses that are part of HPQCD's flavour physics programme on the $N_f=2+1+1$ HISQ ensembles. Related calculations underway include $B$ meson bag parameters, $B\rightarrow \pi l \nu$ form factors at zero recoil and the pion charge radius. 
Including results for charmed decay constants from the MILC collaboration presented at lattice 2013 \cite{claudetalk,dougtalk}, 
we now have precision results for SU(3) breaking ratios with a light, strange, charm or bottom quark on the $N_f=2+1+1$ ensembles. We see for the first time a difference in the ratio as a function of this second quark mass. 


Our result for $w_0$ differs by 2-$\sigma$ from that of BMW \cite{Borsanyi:2012zs} using $N_f=2+1$ HEX smeared Wilson quarks (and also stout smeared staggered). At this conference MILC gave results for $w_0$ on their configurations in good agreement with ours \cite{nathantalk}.  


\vspace{2mm}
{\small \noindent{{\bf Acknowledgements}} We are grateful to the MILC collaboration for the use of their 
gauge configurations. 
We have used the MILC code for some of our propagator 
calculations.  
The results described here were obtained using the Darwin Supercomputer 
of the University of Cambridge High Performance 
Computing Service as part of STFC's DiRAC facility. 
This work was funded by NSF, DoE, the Royal Society, the Wolfson Foundation, SFB-TR 55 and STFC.

\bibliographystyle{hieeetr}
\bibstyle{hieeetr}
\bibliography{rjd_hpqcd_apr2013,rjd_mypapers_apr2013,rjd_milc_apr2013,rjd_bib_apr2013}

\end{document}